\documentclass[fleqn,twoside]{article}%
\usepackage{amsmath}
\usepackage{amsfonts}
\usepackage{amssymb}
\usepackage{graphicx}
\usepackage{cite}
\def\be{\begin{equation}}
\def\ee{\end{equation}}
\def\bi{\bibitem}
\begin{document}
\title{Homogeneous Anisotropic Cosmological Models with Viscous Fluid and Magnetic Field.}
\author{A. Banerjee$^1$, and Abhik Kumar Sanyal$^2$}
\maketitle
\noindent
\begin{center}
\noindent
Department of Physics, Jadavpur University, Calcutta 700 032, India.\\
\end{center}
\footnotetext{\noindent
Electronic address:\\
\noindent
$^1$ asit@juphys.ernet.in\\
$^2$ sanyal\_ ak@yahoo.com;\\
$^2$ Present address: Dept. of Physics, Jangipur College, India - 742213.}
\noindent
\abstract{The paper presents some exact solutions of Bianchi types I, III and Kantowski- Sachs cosmological models consisting of a dissipative fluid along with an axial magnetic field. A barytropic equation of state $(p = \epsilon \rho)$, together with a pair of linear relations between the matter density $(\rho)$, the shear scalar $(\sigma)$, and the expansion scalar $(\theta)$ have been assumed for simplicity. The solutions are basically of two different types, one for the Bianchi-I and the other for Bianchi-III and Kantowski-Sachs type. The presence of the magnetic field, however, does not change the fundamental nature of the initial singularity.}

\maketitle
\flushbottom
\section{Introduction:}

The role of viscosity as a dissipative mechanism responsible for smoothing out the initial large anisotropies of the universe has been investigated by many authors. It was the suggestion of Misner \cite{1} that the neutrino viscosity acting in the early era might have considerably reduced the
anistropy in black body radiation during the evolution. Objections to Misner's method were later raised by other workers such as Doroskevich, Zeldovich, and Novikov \cite{2}, Stewart \cite{3} and Collins and Stewart \cite{4}. Energy conditions and entropy production etc. were discussed by Woszcayna \cite{5} and by Caderni and Fabri \cite{6}. Apart from these qualitative discussions, suitable cosmological models with viscous fluid were also constructed and discussed by Murphy \cite{7}, Banerjee and Santos \cite{8, 9}, Banerjee, Duttachoudhury, and Sanyal \cite{10}, Dunn and Tupper \cite{11}, Coley and Tupper \cite{12}, and others. While there are varieties of homogeneous anistropic cosmological solutions corresponding to perfect fluid, there are not yet many with viscosity terms included in the energy stress tensor. The situation becomes rather more complicated in this case. Since the number of unknowns is now greater, one has to assume more relations between them to obtain exact solutions. Belinsky and Khalatnikov \cite{13} studied the behaviour of anisotropic models containing viscous fluid in the asymptotic limits assuming the viscosity coefficients to be
power functions of the matter density. It was found that the introduction of viscosity terms in the energy momentum tensor of the fluid considerably
change the nature of the initial singularity.\\

We investigate the three classes of spatially homogeneous axially symmetric space-times that contain an imperfect fluid characterized by both bulk and shear viscosity and also having a magnetic field in the axial direction. The corresponding ideal fluid solutions were obtained by many and reviewed by Vajk and Eltgroth \cite{14} with references therein. In our work we assume that the fluid has the property that the ratio of the rate of shear $\sigma$ to the volume expansion $\theta$ is a constant. This assumption was made \cite{15} for a class of solutions with ideal fluid. Another assumption is that the dynamical importance of fluid density does not vary during the evolution. The ratio ${\rho\over \theta^2}$ is very low in most of the homogeneous cosmological models, particularly in the early era [4]. The perfect fluid solutions for the Bianchi-II model with both ${\rho\over \theta^2}$ and ${\sigma^2\over \theta^2}$ constants were first given by Collins and Stewart \cite{14} and later generalized by Banerjee, Duttachaudhury, and Sanyal \cite{16} by introducing viscosity. In all three classes of solutions under consideration it is assumed that there is a magnetic field in the axial direction. These three classes are Bianchi-I, Kantowski-Sachs, and Bianchi-III corresponding to spatial curvature being zero, positive, and negative, respectively. These metric forms were first investigated by Thorne \cite{17}. The fluid in our case obeys barotropic equation of state and the viscosity coefficients are found to be proportional to $\rho^{1\over 2}$ for the spatially flat model, while for nonzero curvature there is no such simple relation. Although the existence of magnetic field influences the metric, it does not alter the basic nature of the initial singularity which occurs at finite past when the energy conditions of Hawking and Penrose are satisfied. \\

In the present work we assume that the fluid flow representing the average motion of matter in the universe is expanding, which is reasonable for a physically realistic model. In the following section, we write the field equations and arrange them properly. In section 3, we attempt exact solution of the field equations under certain physically reasonable assumptions. Case by case study of the solutions have been performed in section 4.

\section{Field equations:}

The metric for the homogeneous anisotropic axially symmetric space-time \cite{14} is

\be\label{2.1} \begin{split} & dS^2 = -dt^2 + e^{2\alpha(t)}dx^2 + e^{2\beta(t)} d\Omega^2,\hspace{0.22 in}\mathrm{where,}\\&
d\Omega^2 = d\theta^2 + sin^2\theta~d\phi^2,\hspace{0.92 in} \mathrm{Kantowski-Sachs,~positive~spatial~curvature},\\&
d\Omega^2 =  dY^2 + dZ^2,\hspace{1.2 in}\mathrm{Bianchi-I,~ zero~ spatial~ curvature}, \\&
d\Omega^2 = d\theta^2 sinh^2\theta~d\phi^2, \hspace{1.0 in}\mathrm{Bianchi-III,~negative~spatial~curvature}.\end{split}\ee
The energy momentum tensor for a viscous fluid along with an axial magnetic field is:
\be\label{2.2} T_{\mu\nu}+E_{\mu\nu} = (\rho+ \bar p)v_\mu v_\nu + \bar p g_{\mu\nu} -\eta U_{\mu\nu} + {1\over 4\pi}\left({F_\mu}^\alpha F_{\nu\alpha} - {1\over 4}F_{\alpha\beta}F^{\alpha\beta}g_{\mu\nu}\right).\ee
The magnetic field is in the $x$ direction and $F_{\alpha\beta}$ is the electromagnetic field tensor, while
\be\label{2.3}\bar p = p - \left(\zeta - {2\over 3}\eta\right){v^\alpha}_{;\alpha}.\ee
\be\label{2.4} v_\mu v^\mu = -1,\hspace{0.7 in} U_{\alpha\beta} = v_{\alpha;\beta} + v_{\beta;\alpha} + v_{\alpha}v^a v_{\beta;a} + v_{\beta}v^a v_{\alpha;a} \ee
\be\label{2.5} {T_0}^0  = -\rho,\hspace{0.78 in}{T_1}^1 = \bar p - 2\eta\dot \alpha, \hspace{0.78 in}{T_2}^2 = {T_3}^3 = \bar p - 2\eta\dot\beta.\ee
Here $\rho$ and $p$ are the matter density and pressure, respectively, $v^\mu$ is the
four velocity, $\zeta$ and $\eta$ are the bulk and shear viscosity coefficients, respectively. All the quantities in the homogeneous models are time-dependent. Since the magnetic field is assumed to be in the $x$ direction, the component $F_{23}$ alone exists. Maxwell's equations $F_{[\mu\nu;\alpha]} = 0$ leads us to the result that $F_{23}$ can at best be a function of $\theta$ alone. The other set of Maxwell's equations $(F^{\mu\nu}\sqrt{-g})_{;\mu} = 0$, then shows that $(F^{23}\sqrt{-g})$ must be independent of $\theta$, which in turn implies
\be\label{2.6} F_{23} = A\psi(\theta),\ee
where $A$ is a constant and $\psi(\theta) = \sin\theta,~1$, or $\sin h\theta$ according to Kantowski-Sachs type, Bianchi-I, or Bianchi-III metric respectively. One can now calculate the energy momentum tensors for the electromagnetic field. These are:
\be\label{2.7} {E_0}^0 = {E_1}^1= -{E_2}^2 = -{E_3}^3 = -\left(A^2\over 8\pi\right)e^{-4\beta}.\ee
Choosing $8\pi G = 1$, Einstein's field equations are now
\be\label{2.8} {R_\mu}^\nu - {1\over 2} {\delta_\mu}^\nu R = - \left({T_\mu}^\nu + {E_\mu}^\nu\right). \ee
Different independent components of \eqref{2.8} are given by

\be\label{2.9} {9\over 2}\left({\dot R^2\over R^2}\right) - {1\over 2}\left(\dot \alpha^2 + 2\dot\beta^2\right)+ k e^{-2\beta} = \rho + \left({A^2\over 8\pi}\right) e^{-4\beta},\ee

\be\label{2.10} 2\ddot\beta + {3\over 2}\left({\dot R\over R}\right)(2\dot\beta - \dot \alpha) + {1\over 2}\left(\dot \alpha^2 + 2\dot\beta^2\right)+ k e^{-2\beta} = -\left[\bar p - 2\eta\dot\alpha - \left({A^2\over 8\pi}\right) e^{-4\beta}\right],\ee

\be\label{2.11} \ddot\alpha + \ddot\beta + {3\over 2}\left({\dot R\over R}\right)\dot \alpha + {1\over 2}\left(\dot \alpha^2 + 2\dot\beta^2\right) = -\left[\bar p - 2\eta\dot\beta - \left({A^2\over 8\pi}\right) e^{-4\beta}\right],\ee
where $k= +1,~ 0,~\mathrm{or}~ -1$ according to whether the model is
Kantowski-Sachs, Bianchi-I, or Bianchi-III, respectively. In the above we have written $R^3$ for $\sqrt{-g}$, so that the expansion scalar $\theta$ and the shear $\sigma^2$ are expressed in the form
\be\label{2.12} \theta = {v^\alpha}_{;\alpha} = \dot\alpha + 2\dot\beta = 3\left(\dot R\over R\right),\ee
and
\be\label{2.13} 2\sigma^2 = \sigma_{\mu\nu}\sigma^{\mu\nu} = \dot \alpha^2 +2\dot\beta^2 -{1\over 3}\theta^2,\ee
where,
\be\label{2.14} \sigma_{\mu\nu} = {1\over 2} U_{\mu\nu} - {1\over 3}\theta\big(g_{\mu\nu} + v_{\mu}v_{\nu}\big).\ee
Now using equations \eqref{2.3}, \eqref{2.12}, and \eqref{2.13}, we can express the field equations \eqref{2.9} to \eqref{2.11} in the following form:
\be\label{2.15} {1\over 3}\theta^2 - \sigma^2 - \rho = \left({A^2\over 8\pi}\right) e^{-4\beta} - k e^{-2\beta},\ee
\be\label{2.16} \ddot\beta + \dot\beta \theta + 2\eta \left(\dot\beta - {\theta\over 3}\right)- {1\over 2}\zeta\theta -{1\over 2}(\rho - p)  -\left({A^2\over 8\pi}\right) e^{-4\beta} + k e^{-2\beta} = 0,\ee
\be\label{2.17}\ddot\beta + \dot\beta \theta + 2\eta \left(\dot\beta - {\theta\over 3}\right)+ \zeta\theta + (\rho - p) -\dot\theta -\theta^2 - k e^{-2\beta} = 0.\ee
In deriving the last two equations we have utilized \eqref{2.15}. The equation of motion ${{T_\mu}^\nu}_{;\nu} = 0$ (${{E_\mu}^\nu}_{;\nu} = 0$ since there is no charge) can now be expressed as
\be\label{2.18} \dot \rho + (\rho + p)\theta - \zeta\theta^2 - 4\eta \sigma^2 = 0. \ee
Again in view of equations \eqref{2.16} and \eqref{2.17} we get
\be\label{2.19} \dot \theta = {3\over 2}(\rho - p) +{3\over 2}\zeta\theta -\theta^2 + \left({A^2\over 8\pi}\right) e^{-4\beta} - 2k e^{-2\beta} = 0\ee
which, in view of \eqref{2.15}, is again
\be\label{2.20} \dot\theta = -{1\over 3}\theta^2 - 2\sigma^2 -{1\over 2}\left[\rho + 3(p - \zeta\theta)+ \left({A^2\over 8\pi}\right) e^{-4\beta} \right],\ee
and may easily be identified as the Raychaudhuri's equation \cite{18}; that is
\be\label{2.21} \dot\theta = -{1\over 3}\theta^2 - 2\sigma^2 + R_{\mu\nu} v^\mu v^\nu.\ee
So, from equations \eqref{2.15}, \eqref{2.18}, and \eqref{2.19}, we get the relation
\be\label{2.22}\begin{split} \left({\rho\over\theta^2}\right)^\textbf{.} &= -\left({\sigma^2\over\theta^2}\right)^\textbf{.} -\left[{1\over \theta^2}\left({A^2\over 8\pi}e^{-4\beta} - ke^{-2\beta} \right)\right]\\& = \left({\sigma^2\over\theta^2}\right)\left(4\eta + 3\zeta + {3(\rho - p)\over \theta}\right) + {1\over \theta^2}\times{A^2\over 8\pi}e^{-4\beta}\left(3\zeta + {\rho - 3p \over \theta}\right) - {k\over \theta^2}e^{-2\beta}\left(3\zeta - {\rho + 3p \over \theta}\right).\end{split}\ee
We obtained the special case of \eqref{2.22} for Bianchi-I metric ($k = 0$) and in the absence of the magnetic field in a previous paper \cite{10}. For the Bianchi-I model, i.e., $k=0$ in general, for matter density $\rho$ greater than $3p$ and positive viscosity coefficients, we have $\left({\rho\over\theta^2}\right)^\textbf{.} > 0$, which means that the dynamical importance of the density increases in the course of time. In the same case, when $\left({\rho\over\theta^2}\right)$=constant, the limits on $\rho$ and $p$ should be $p< \rho < 3p$. In the case for $k= +1$, $A^2 \ne 0$, and $\zeta = 0$, i.e for vanishing bulk viscosity and also $\rho > 3p$, again we have $\left({\rho\over\theta^2}\right)^\textbf{.}> 0$. We give some exact solutions under certain restrictions below.

\section{Exact solutions of the field equations:}

We now introduce three additional equations, the barytropic equation of state and the two others mentioned in the introduction. These are

\be\label{2.23} p = \epsilon \rho,\ee
\be\label{2.24} \rho = C^2 \theta^2,\ee
\be\label{2.25} \sigma^2 = D^2 \theta^2.\ee
We now have six equations viz. the three field equations, \eqref{2.15} to \eqref{2.17},
together with \eqref{2.23} to \eqref{2.25}, to determine the six unknowns $\alpha,~\beta,~\rho,~p,~\eta,~\zeta$. A class of homogeneous solutions where $\left({\sigma^2\over \theta^2}\right)$= constant was previously discussed \cite{15}. The present-day upper limit of this ratio, however, is argued to be very small compared to unity [18]. It may be mentioned further in connection with relation \eqref{2.24}, that it is indeed true for the Einstein-deSitter universe which is isotropic and has matter density equal to the critical value which, if exceeded, indicates that the universe is closed.\\

Using equations \eqref{2.23} to \eqref{2.24}, field equations \eqref{2.15} to \eqref{2.17} reduce to the forms
\be\label{2.26} \left({1\over 3} -C^2 - D^2\right)\theta^2 = {A^2\over 8\pi} e^{-4\beta} - ke^{-2\beta}, \ee
\be\label{2.27} \ddot \beta + \dot\beta\theta + 2\eta\left(\dot\beta -{\theta\over 3}\right) -{1\over 2}\zeta\theta - {1-\epsilon\over 2}C^2\theta^2 - {A^2\over 8\pi} e^{-4\beta} +ke^{-2\beta} = 0, \ee
\be\label{2.28} \ddot \beta + \dot\beta\theta + 2\eta\left(\dot\beta -{\theta\over 3}\right) + \zeta\theta - \left[1 - (1-\epsilon)C^2\right]\theta^2 -\dot\theta - ke^{-2\beta} = 0. \ee
The bulk viscosity expression can be obtained from \eqref{2.19} or (20) as
\be\label{2.29a} \begin{split}\zeta &= {2\over 3}\left[{\dot\theta\over \theta } + \left({2\over 3} + {3\epsilon -1\over 2}C^2 + D^2\right)\theta\right] +{2\over 3}\times {k\over \theta}e^{-2\theta}\\& = {2\over 3}\left[{\dot\theta\over \theta } + \left({1\over 3} + {3\epsilon + 1\over 2}C^2 + 2D^2\right)\theta + {1\over \theta}\times {A^2\over 8\pi}e^{-4\theta}\right].\end{split}\ee
The above relations can also be obtained from \eqref{2.27} and \eqref{2.28} using \eqref{2.26}. The shear viscosity can now be obtained from \eqref{2.18} and is expressed in the form
\be\label{2.29b} \eta =  {1\over 2 D^2}\left[{1\over 3}\left(2C^2 - D^2 - {2\over 3}\right)\theta - \left({1\over 3} - C^2\right){\dot\theta\over \theta} - {k\over 3\theta}e^{-2\theta}\right].\ee
Solutions of the above set of equations are not difficult to obtain. From
\eqref{2.12} and \eqref{2.13} we get
\be\label{2.30}\dot \alpha = \theta - 2\dot\beta, \ee
and
\be\label{2.31} 2\sigma^2 = \theta^2 +4\beta^2 - 4\dot\beta \theta + 2\dot\beta^2 - {1\over 3}\theta^2 = {2\over 3}\theta^2 + 6\dot\beta^2 - 4\dot\beta \theta,\ee
which again in view of \eqref{2.25} reduces to a quadratic equation in $\dot\beta$, that is
\be\label{2.31a} 3\dot\beta^2 - 2\dot\beta\theta +\left({1\over 3} - D^2\right)\theta^2 = 0, \ee
so that,
\be\label{2.32} \dot \beta = {1\over 3}\Big[\theta \pm \sqrt{3D^2 \theta^2}\Big] = \left[{1\over 3} \pm {D\over \sqrt{3}}\right]\theta.\ee

\section{Solutions of the field equations, case study:}
\subsection{Case-I: $k = 0$.}
This defines the Bianchi-I space-time. We consider the equation (2b)
which can now be written as
\be\label{2.33} e^{-4\beta}  = P^2 \dot \beta^2,\ee
where the constant $P^2$ is written for ${8\pi \over A^2 h^2}\big({1\over 3} - C^2 - D^2\big)$ , where
\be\label{2.33a} h = {1\over 3} \pm \Big({D\over \sqrt{3}}\Big).\ee
Now the integration of the equation \eqref{2.33} yields
\be\label{2.34} e^{2\beta} = \Big({2\over P}\Big)t + q,\ee
$q$ being an arbitrary integration constant. The relation \eqref{2.32} now yields
\be\label{2.35} \theta = \big[2ht + m\big]^{-1},\ee
where the constant $m$ is written for $hpq$. It can be found easily that both the viscosity coefficients $\zeta$ and $\eta$ are now proportional to the expansion scalar $\theta$, which in turn is proportional to $\rho^{1\over 2}$. The matter density $\rho$ and the shear $\sigma^2$ can be expressed explicitly since the solution for $\theta$ is unknown. Writing $e^{2\beta}$ in \eqref{2.34} in the form,
\be\label{2.36} e^{2\beta} = {1\over h P} (2ht + m),\ee
the solution for $\alpha$ can be obtained by integrating \eqref{2.30} to yield
\be\label{2.37} e^{2\alpha} = \alpha_0\big[2h t + m\big]^{1-2h\over h},\ee
where, $\alpha_0$ is a constant. When the fluid has only shear viscosity and negligible bulk viscosity we may assume $\zeta = 0$, which leads, in view of \eqref{2.29a}, to the following relations valid for an expanding model ($\theta > 0$) (which is a reasonable model universe)
\be\label{2.38} -2h + {2\over 3} + D^2 + {3\epsilon - 1\over 2}C^2 = 0.\ee
Further, \eqref{2.26} indicates that
\be\label{2.39} C^2 + D^2 -{1\over 3} < 0.\ee
Combining \eqref{2.38} and \eqref{2.39} we obtain
\be\label{2.39a} C^2 + D^2 -{1\over 3} < -2h + {2\over 3} + D^2 + {3\epsilon - 1\over 2}C^2,\ee
or,
\be\label{2.40} (1-2h) > \left[{3\over 2}(1 - \epsilon)\right]C^2.\ee
Since the maximum value of $\epsilon$ can only be unity in the Zeldovich limit,
we have $(1 - 2h) > 0$. As a consequence, when $[2ht + m]$ approaches zero
we have both $e^{2\alpha}$ and $e^{2\beta}$ vanish, giving rise to a point singularity with $\theta\rightarrow \infty$. On the other hand, if $\zeta \ne 0$, it is possible to choose the constants in such a way that $(1-2h) < 0$. In that case, $e^{2\alpha}\rightarrow \infty$ while $e^{2\beta} \rightarrow 0$ in the initial epoch, giving rise to a line singularity along $x$ direction, that is along the direction of the magnetic field. At the singularity, however, along with the expansion scalar $\theta$ the other quantities such as $\rho,~ \sigma^2$, and the magnetic field increase indefinitely. The total proper volume is proportional to $e^{\alpha + 2\beta}$; that is, $R^3$ depends directly on $(2ht+m)^{1\over 2h}$. Thus when $e^{2\beta} \rightarrow 0$, one has the proper volume vanishingly small, and on the other hand, when $t \rightarrow \infty$
the proper volume increases to infinitely large magnitude. The model in this limit, however, takes the shape of a disk, because $e^{2\alpha} \rightarrow 0$ but $e^{2\beta} \rightarrow \infty$.\\

\noindent
\textbf{Energy Conditions:}
The shear viscosity expression \eqref{2.29b} in the case when \eqref{2.35} is used reduces to
\be\label{2.41}\begin{split} \eta &= {\theta\over 2D^2}\left[{1\over 3} \left(2C^2 - D^2 -{2\over 3}\right) + \left({1\over 3 } - C^2\right)\left({2\over 3}\pm {2\over \sqrt 3} D\right)\right]\\&
={\theta\over 2D^2}\left[-{D^2\over 3} \pm {2D\over \sqrt 3}\left({1\over 3} -C^2\right)\right]. \end{split}\ee
Since we know that, in view of \eqref{2.26}, $C^2 < {1\over3}$ for an expanding model under consideration, $\eta > 0$ only if we choose $\pm D$ greater than $0$. Let us assume $D$ positive; then, for a physically well-behaved fluid, $2h ={2\over 3} + ({2D\over \sqrt 3})$. Thus from \eqref{2.21} using \eqref{2.32} we have
\be\label{2.50}\begin{split} - R_{\mu\nu}v^\mu v^\nu &= -{1\over 3} \theta^2 - 2\sigma^2 - \dot\theta = \left[-{1\over 3} - 2D^2 + \left({2\over 3} + {2D\over \sqrt 3}\right)\right]\theta^2\\&
=\left[{1\over 3} + 2D\left({1\over \sqrt 3} - D\right)\right]\theta^2. \end{split}\ee
Again, since $D < {1\over \sqrt 3}$, it is possible to conclude that $-R_{\mu\nu} v^\mu v^\nu > 0$, so that Hawking's energy condition is satisfied and the initial zero volume singularity is unavoidable.

\subsection{Case-II: $k = \pm 1$.}

These define Kantowski-Sachs and Bianchi-III models corresponding to $k= +1$ and $k= -1$, respectively. In this case equation \eqref{2.26} can be written as

\be\label{2.51} e^{-4\beta} + ne^{-2\beta} = P \dot \beta^2,\ee
where, $P$ as before equals $\left({8\pi\over A^2 h^2}\right)({1\over 3} -C^2 - D^2)$ and $n = -{8\pi k\over A^2}$. Integrating equation \eqref{2.51} we get,

\be\label{2.52} e^{2\beta} = \left[{n\over P}(t + h^2)^2 - {1\over n}\right].\ee
Writing constants as
\be\label{2.52a} f = {n\over P} = -\left[{k\left({1\over 3} \pm {D\over \sqrt 3}\right)^2\over {1\over 3}-c^2 -D^2}\right] = -\left[{k h^2 \over {1\over 3}-c^2 -D^2}\right],\;\;\;\mathrm{and}\;\;\;g = {1\over n} = - {A^2\over 8\pi k},\ee
and with a suitable time transformation $T= (t+h)$ the solution \eqref{2.52} is
expressed as
\be\label{2.53} e^{2\beta} = \left(fT^2 - g\right).\ee
The relation \eqref{2.32} leads to
\be\label{2.54} \theta = {fT\over h\left(fT^2 - g\right)}.\ee
Using \eqref{2.53} and \eqref{2.54} one can integrate equation \eqref{2.30} to obtain
\be\label{2.55} e^{2\alpha} \propto \left(fT^2 - g\right)^{1-2h\over h}.\ee
The above solutions are found to satisfy field equations \eqref{2.26} to \eqref{2.28} when substituted directly in them. The procedure, however, is lengthy to some extent. As in the previous case, let us now consider the situation when the bulk viscosity is negligible so that $\zeta = 0$, which leads to the relation (for $k = \pm 1$)
\be\label{2.56} {k\over \theta} e^{-2\beta} = {\dot\theta\over \theta} + \left({2\over 3}+{3\epsilon - 1\over 2} C^2 + D^2\right)\theta.\ee
Substituting equation \eqref{2.56} in expression \eqref{2.29a} for $\eta$, we finally get
\be\label{2.57}\eta = {C^2 \theta\over 2 D^2}\left[{\dot\theta\over \theta^2} + {1\over 2}(1 + \epsilon)\right].\ee
Using the solution for $\theta$ given in equation \eqref{2.54}, one can express $\eta$ also in
the following form
\be\label{2.58} \eta = {C^2 \theta\over 2 D^2}\left[-h - {hg\over fT^2} + {1\over 2}(1 + \epsilon)\right].\ee
In general, at singularity $e^{2\beta} \rightarrow 0$, we have an infinitely large magnetic field. For $k = -1$, that is, for Bianchi-lII model; we have from \eqref{2.26}, $\big({1\over 3} -C^2 - D^2\big) > 0$ which means that both $f$ and $g$ are positive, and as a consequence we must have $D < {1\over \sqrt 3}$. The term $h = \big({1\over 3}\pm {D\over \sqrt 3}\big)$ is positive definite independent of the sign chosen. For such a model there occurs three different situations corresponding to the constant $2h$ less than, equal to, or greater than unity. For $2h < 1$ we have $\theta \rightarrow \infty$ and $e^{2\alpha} \rightarrow 0$, $e^{2\beta} \rightarrow 0$ as $T \rightarrow \big({g\over f}\big)^{l\over 2}$. This represents a point singularity where $\rho$ and $\sigma^2$ both become infinitely large. For $2h > 1$, however, the situation is completely different. In this case, $e^{2\alpha} \rightarrow \infty$, when $e^{2\beta} \rightarrow 0$. The expansion scalar $\theta \rightarrow \infty$ but the singularity is line type. In the third case, $2h = 1$, we have $e^{2\alpha} =$ constant, so that the singularity is rod-like. In the above discussions the singularity exists as $T \rightarrow \big({g\over f}\big)^{l\over 2}$, so that ${fT^2\over g} = 1$ and we have for the shear viscosity,
\be\label{2,59}\eta = {C^2 \theta\over 2 D^2}\left[-2h  + {1\over 2}(1 + \epsilon)\right],\ee
from equation \eqref{2.58}. So, for an expanding model $\theta > 0$ with suitable values of the constant, parameters $\eta$ can remain positive at this instant; once it is greater than zero it remains so for any subsequent instant with the increase of $T$.\\

On the other hand, the situation may be different for the Kantowski-Sachs model ($k= +1$). In this case $g = -{A^2\over 8\pi}$ and for the reasonable model of the universe we have $\theta > 0$, while $C^2$ and $D^2$ are also very small. These physical conditions demand ${1\over 3} - C^2 - D^2 > 0$ so that $f < 0$. The proper volume is given by $R^3 = e^{\alpha+2\beta}$, which is now $R^3 \sim (|g|- |f| T^2)^ {1\over 2h}$ and $\theta \sim -\big[{|f|T\over h}(|g|- |f| T^2)\big]$. Here $|f|$ and $|g|$ represent the magnitudes of $f$ and $g$, respectively. The expanding phase of the model is described only for negative values of $T$ here, so that when $T \rightarrow -|{g\over f}|^{1\over 2}$,
$\theta\rightarrow +\infty$ and $R^3 \rightarrow 0$. Again when $T \rightarrow 0$, we have $\theta \rightarrow 0$ and $R^3 \rightarrow |g|^{1\over 2h}$. From the above analysis it is clear that there is an upper limit for the expansion, and so an expanding model starting from a singularity turns back after reaching a finite proper volume, finally approaching the singularity of zero volume again. At the singularity the density and pressure of the fluid content attain infinitely large magnitude. In the Kantowski-Sachs special case with positive curvature, the zero volume singularity $R^3 \rightarrow 0$ occurs at $T=-|{g\over f}|^{1\over 2}$, or in other words, when ${hg\over fT^2}= h$. One notes in this case that the shear viscosity $\eta$ may be a finite positive quantity with suitable magnitudes of the constant parameters. But on the other hand, the same parameter attains a large negative magnitude as $T$ approaches zero value in the future, which of course introduces undesirable features into the behavior of the fluid.\\

\noindent
\textbf{Energy Conditions:}\\
For $k \ne 0$ one can calculate

\be\label{2.60} -R_{\mu\nu} v^\mu v^\nu = -{1\over 3}\theta^2 - 2\sigma^2 -\dot\theta = \left[-{1\over 3} - 2D^2 + h + {hg\over fT^2}\right]\theta^2.\ee
Since in all the above cases ${hg\over f} > 0$, the last term is always greater than zero. It is therefore possible to adjust the other constants in such a way
that the energy condition $-R_{\mu\nu} v^\mu v^\nu > 0$ is satisfied. It is, however, a straightforward matter to calculate $\dot\theta$ from expression \eqref{2.54}. We have

\be\label{2.61} \dot \theta = \left[- h - {hg\over fT^2}\right]\theta^2.\ee
So we observe that so long as $h > 0$ and ${g\over f} > 0,~ \dot\theta < 0$, which means that the collapse is unavoidable.\\

\noindent
\textbf{Condition for Entropy Change:}\\
The time derivative of the entropy density can be given \cite{13} as

\be\label{2.62} {\dot\Sigma\over \Sigma} = \dot\rho(\rho + p),\ee
where $\Sigma$ is the entropy density. Defining the total entropy by $S = R^3\Sigma$ one can find, in view of equations \eqref{2.18} and \eqref{2.23}

\be\label{2.63} {\dot S\over S} = {\zeta + 4\eta \left({\sigma^2 \over \theta^2}\right)\over (1+ \epsilon)\left({\rho \over \theta^2}\right)}.\ee
Since ${\sigma^2 \over \theta^2} > 0$ and ${\rho \over \theta^2} > 0$ in view of equations \eqref{2.24} and \eqref{2.25}, ${\dot S\over S} > 0$ as long as viscosity coefficients are positive. Hence $\dot S > 0$, which implies that
the total entropy always increases with the change of proper time, irrespective
of an expanding or contracting model.\\

\noindent
\textbf{Acknowledgement:}
Thanks are due to U.G.C. (India) for financial support and to the referee for bringing to our attention some points overlooked earlier.

\end{document}